\documentclass[twocolumn]{revtex4}
\usepackage{amssymb}
\usepackage{latexsym}
\usepackage{float}
\usepackage{epstopdf}

\begin{document}

\title{The extended uncertainty principle inspires the R\'{e}nyi entropy}
\author{H. Moradpour$^{1}$\footnote{h.moradpour@riaam.ac.ir}, C. Corda$^{2}$\footnote{cordac.galilei@gmail.com}, A. H. Ziaie$^{1}$\footnote{ah.ziaie@riaam.ac.ir}, S. Ghaffari$^{1}$\footnote{sh.ghaffari@riaam.ac.ir}}
\address{$^{1}$ Research Institute for Astronomy and Astrophysics of Maragha (RIAAM), Maragha 55134-441, Iran\\ $^2$ International Institute for Applicable Mathematics and
Information Sciences, B. M. Birla Science Centre, Adarshnagar,
Hyderabad 500063, India}

\begin{abstract}
We use the extended uncertainty principle (EUP) in order to obtain
the R\'{e}nyi entropy for a black hole (BH). The result implies
that the non-extensivity parameter, appeared in the R\'{e}nyi
entropy formalism, may be evaluated from the considerations which
lead to EUP. It is also shown that, for excited BHs, the R\'{e}nyi
entropy is a function of the BH principal quantum number, i.e. the
BH quantum excited state. Temperature and heat capacity of the
excited BHs are also investigated addressing two phases while only
one of them can be stable. At this situation, whereas entropy is
vanished, temperature may take a non-zero positive minimum value,
depending on the value of the non-extensivity parameter. The
evaporation time of excited BH has also been studied.
\end{abstract}

\maketitle

\section{Introduction}

In one hand, uncertainty principle inspires Bekenstein entropy,
and indeed, various generalized uncertainty principles (GUP) add
different modifications to Bekenstein entropy
\cite{Bek,maju1,ahmed1,ahmed2,ahmed3}. On the other hand,
Bekenstein entropy is a non-extensive entropy measure, a property
which motivates some physicists to consider it as a suitable
candidate for the Tsallis entropy in calculating the R\'{e}nyi
entropy \cite{kom,prd,rhd,non21,prdinf}. In fact, due to the
long-range nature of gravity \cite{pla}, the use of generalized
entropy formalisms such as those introduced by Tsallis \cite{tsa}
and R\'{e}nyi \cite{ren} has recently been taken into
consideration \cite{kom,prd,rhd,non21,prdinf,Tavayef,smm}.

A long-range interacting system with $W$ discrete states while each
state $i$ has probability $P_i$ may follow a power probability
distribution instead of the ordinary distribution
\cite{pla,tsa,ren}. Working with
$G=c=k_{B}=\hbar=\frac{1}{4\pi\epsilon_{0}}=1$ (Planck units), where
$k_B$ denotes the Boltzmann constant, the Tsallis entropy of such
system is defined as \cite{tsa}

\begin{eqnarray}\label{reyn01}
S_T=\frac{1}{1-q}\sum_{i=1}^{W}(P_i^q-P_i),
\end{eqnarray}

\noindent where $q$ is an unknown parameter \cite{pla}. It is
worthwhile mentioning that one can reach $S_T=\frac{A}{4}$, where
$A$ denotes the horizon of system (boundary), by applying the
Tsallis entropy definition~(\ref{reyn01}) to the gravitational
systems \cite{5}. This result is in agreement with the cosmological
studies in which authors assumed $S_T=\frac{A}{4}$ in calculating
R\'{e}nyi entropy \cite{kom,prd,rhd,non21,prdinf} written as
\cite{pla}

\begin{eqnarray}\label{reyn1}
\mathcal{S}=\frac{1}{\delta}\ln(1+\delta S_T),
\end{eqnarray}

\noindent in which $\delta\equiv1-q$. In addition to the successes
of this entropy in describing cosmos \cite{kom,prd,rhd,prdinf}, it
can also be combined with the Verlinde's hypothesis \cite{ver} to
give us a theoretical basis for the MOND theory and its
modifications \cite{non21}. In this paper, we are going to show
that EUP can also lead to the emergence of R\'{e}nyi entropy. In
addition, relation between $\delta$ (and thus $q$) and the quantum
mechanical parameter appeared in EUP is also derived. After
getting the mentioned aim in the next section, we study some
thermodynamic properties of excited BHs meeting R\'{e}nyi entropy
in the third section. The last section is devoted to summary and
concluding remarks.

\section{From EUP to R\'{e}nyi entropy}

In the framework of the high energy physics, such as quantum
gravity, various GUP and EUP are derived
\cite{prdgup,eup1,eup2,eup3,Bek}, which can generally be written
as \cite{prdgup}

\begin{eqnarray}\label{gupk}
\Delta x\Delta p\geq[1+\beta(\Delta x)^2+\eta(\Delta p)^2+\gamma].
\end{eqnarray}

\noindent Here, $\gamma$ is positive and depends on the
expectation values of $p$ and $x$ \cite{prdgup,norazi}. Bearing
the fact that the minimum uncertainty is obtainable for $\gamma=0$
in mind \cite{norazi}, we consider the $\eta=\gamma=0$ case which
leads to EUP written as \cite{prdgup,eup1,eup2,eup3,Bek,prd0}

\begin{eqnarray}\label{gup1}
\Delta x\Delta p\geq[1+\frac{\beta\pi}{4}(\Delta x)^2],
\end{eqnarray}

\noindent where $\beta$ is positive and independent of the values
of $\Delta x$ and $\Delta p$ \cite{prdgup}. The non-zero minimal
values of $\Delta x$ and $\Delta p$, called $\Delta x_0$ and
$\Delta p_0$, respectively, are obtainable whenever $\beta>0$
\cite{prdgup}. The above EUP affects the early universe
thermodynamics \cite{norazi}, and in general, there are deep
connections between EUP and GUP and $i$) the dispersion relation
\cite{dis1,dis2}, $ii$) the Chandrasekhar and Jeans limits and the
dark energy problem \cite{prd0,mnras}. More studies on the
outcomes of employing GUP in various branches of physics can also
be found in
\cite{Bek,maju1,ahmed1,ahmed2,ahmed3,deltag,deltax1,deltax2,deltax3,deltae2,deltae1,pedram}.

Whenever the EUP~(\ref{gup1}) is valid, one can write $\Delta
E\approx\Delta p$ for the uncertainty of the particle energy
($\Delta E$) \cite{ahmed1,ahmed2,ahmed3,deltax3,deltae2,deltae1},
leading to

\begin{eqnarray}\label{gup2}
\Delta E\geq\frac{1}{\Delta x}[1+\frac{\beta\pi}{4}(\Delta
x)^2].
\end{eqnarray}

\noindent In BH physics, whenever a BH with area
$A$ absorbs or emits a quantum particle with energy $E$ and size
$R$, then the changes in the BH area  follows the
$\Delta A\geq8\pi ER$ relation \cite{deltax1,deltax2,deltax3}. Since
the size of a quantum particle cannot be less than the uncertainty in
its position \cite{deltaamin,deltax2,deltax3}, one reaches $\Delta
A_{min}\geq8\pi E \Delta x$ for a quantum particle
\cite{deltax2,deltax3,ahmed3}. Combining this result with
Eq.~(\ref{gup2}), we reach at

\begin{eqnarray}\label{gup3}
\Delta A_{min}\geq8\pi[1+\frac{\beta\pi}{4}(\Delta x)^2].
\end{eqnarray}

\noindent As it has been argued in
Refs.~\cite{deltax1,deltax2,deltax3,ahmed3}, one can write $(\Delta
x)^2\approx\frac{A}{\pi}$ and insert it in Eq.~(\ref{gup3}) to
obtain

\begin{eqnarray}\label{gup4}
\Delta A_{min}\simeq8\pi\lambda[1+\frac{\beta}{4} A],
\end{eqnarray}

\noindent where $\lambda$ is an unknown coefficient fixed later
\cite{ahmed3}.

Therefore, $\Delta A_{min}$ is the minimum changes in the boundary
$A$ whenever EUP~(\ref{gup1}) is valid. It is also obvious to
assume that the corresponding entropy changes is also minimum and
equal to one bit of information $\Delta S_{min}=b=\ln2$
\cite{ver,deltaamin,ahmed3}. The above argument motivates us to
write

\begin{eqnarray}\label{ent1}
\frac{dS}{dA}=\frac{\Delta S_{min}}{\Delta
A_{min}}=\frac{b}{8\pi\lambda[1+\frac{\beta}{4} A]},
\end{eqnarray}

\noindent leading to

\begin{eqnarray}\label{ent2}
S=\frac{b}{2\pi\lambda\beta}\ln[1+\frac{\beta}{4} A].
\end{eqnarray}

\noindent In the limit of
$\beta\rightarrow0$, the Bekenstein entropy ($\frac{A}{4}$) should
be recovered \cite{Bek,maju1,ahmed1,ahmed2,ahmed3,deltaamin,serd} which yields

\begin{eqnarray}\label{ent3}
\lambda=\frac{b}{2\pi},
\end{eqnarray}

\noindent whereby we get

\begin{eqnarray}\label{ent4}
S=\mathcal{S},\ \ \delta=\beta.
\end{eqnarray}

\noindent Thus, one can realize that $i$) EUP may allow us to
employ the R\'{e}nyi entropy, and in this situation, $ii$) the EUP
parameter $\beta$ determines the value of the non-extensivity
parameter $\delta$. Finally, it is also worthwhile mentioning that
the value of $\lambda$ obtained in Eq.~(\ref{ent3}) is the same as
that of the previous work by other authors \cite{ahmed3} in which
the $\beta=\gamma=0$ case has been studied.

\section{Applications to the excited BHs}

For excited BHs, i.e. the BHs which emitted a large amount of
Hawking quanta, the recent Bohr-like approach to BH quantum physics
in \cite{Bohr,Bohr2,Bohr3} permits to write the Bekenstein entropy
in terms of the BH quantum level as \cite{Bohr3}

\begin{equation}\label{0ent}
\frac{A}{4}=4\pi\left(M^{2}-\frac{n}{2}\right),
\end{equation}

\noindent where $M$ is original BH mass and ${n}$ is the BH
principal quantum number if the BH is seen as ${gravitational}$
${atom}$, see e.g., \cite{Bohr,Bohr2,Bohr3}. We indeed recall that, the intuitive but general belief~\cite{Bohr,Bohr2,Bohr3}: the
BHs result in highly excited states representing both the
${Hydrogen}$ ${atom}$ and the ${quasi-thermal}$ emission in quantum
gravity, has been shown to be correct, because the Schwarzschild BH
results in somewhat similar to the historical semi-classical
hydrogen atom introduced by Bohr in 1913, see
\cite{Bohr,Bohr2,Bohr3} for more details. Thus, by using Eqs. (2) and
(11), the  R\'{e}nyi entropy becomes function of the BH principal
quantum number, i.e., the BH excited state, given as

\begin{equation}\label{ent0}
S=\beta^{-1}\ln\left[1+4\beta\pi\left(M^{2}-\frac{n}{2}\right)\right].
\end{equation}

\noindent Accepting the $E=M$ relation and bearing the
$T=\frac{\partial E}{\partial S}$ relation in mind, one reaches

\begin{equation}\label{tem}
T=\frac{1+4\beta\pi\left(M^{2}-\frac{n}{2}\right)}{8\pi M},
\end{equation}

\noindent for temperature of the Hawking radiation in this
formalism. As a check, the temperature $T_B=\frac{1}{8\pi M}$,
obtained by using the Bekenstein entropy, is also recovered at the
appropriate limit $\beta\rightarrow0$. In this manner, for
$n\rightarrow n_{max}=2M^2$ \cite{Bohr3}, we have $S\rightarrow0$
and $T\rightarrow T_B$ independent of the value of $\beta$.
Moreover, the heat capacity evaluated as

\begin{equation}\label{c}
C=\frac{\partial M}{\partial T}=\frac{8\pi
M^2}{2\beta\pi\left(2M^{2}+n\right)-1},
\end{equation}

\noindent includes a singularity at $n=n_{max}$ whenever
$\beta=\frac{1}{8\pi M^2}\equiv\beta_0$. For this critical value of
$n$, we have $C>0$ ($C<0$) for $\beta>\beta_0$ ($\beta<\beta_0$),
and in neighboring of this point one can write
$C\sim\frac{1}{\beta-\beta_0}$. This means that the $\beta>\beta_0$
($\beta<\beta_0$) phase can be stable (unstable)
\cite{nonex,smbh,callen}. For the critical value $\beta_0$, one can
write~(\ref{tem}) as

\begin{equation}\label{tem1}
T=\frac{T_B}{2}[3-\frac{n}{2M^2}],
\end{equation}

\noindent indicating $T>T_B$ for $n<n_{max}$ which means that $T_B$
is the minimum possible temperature at this situation. More studies
on the non-excited BHs ($n=0$) as well as their thermodynamics in
the framework of the R\'{e}nyi entropy can also be found in
\cite{nonex,nonex1,nonex2}.

Now, let us look at the radiation of excited BH as a black body
radiation, and write \cite{smbh,nonex1}

\begin{equation}
\frac{dM}{dt}=-16\pi M^2\sigma T^4(M),
\end{equation}

\noindent where $\sigma$ denotes the Stefan-Boltzman constant, and
additionally, we assumed that the $r=2M$ relation is still valid
\cite{Bohr3}. Therefore, by using Eq.~(\ref{tem}), the time that a
BH needs to lose its mass $M$ can be evaluated as

\begin{equation}\label{t1}
t=-\frac{4(4\pi)^3}{\sigma}\int_{M}^{0}\frac{M^2dM}{(1+4\beta\pi\left(M^{2}-\frac{n}{2}\right))^4},
\end{equation}

\noindent leading to

\begin{eqnarray}\label{t2}
&&\tilde{t}=\frac{3\tan^{-1}[\frac{\sqrt{8\pi\beta}M}{\sqrt{2-4\pi\beta
n}}]}{\sqrt{1-2\pi\beta n}}+\frac{1}{(4\pi\beta(n-2M^2)-2)^3}\bigg(\nonumber\\
&&2\sqrt{4\pi\beta}M\Big((4\pi\beta)^2[3n^2+16nM^2-12M^4]+12\nonumber\\&&-16\pi\beta(3n+8M^2)\Big)\bigg),
\end{eqnarray}

\noindent where $\tilde{t}\equiv3\sigma
t(\frac{\beta}{4\pi})^{\frac{3}{2}}$. Whenever $\beta=0$ or even
$n=n_{max}$, one obtains

\begin{eqnarray}\label{t3}
t=\frac{4(4\pi)^3}{3\sigma}M^3,
\end{eqnarray}

\noindent the evaporation time of a Schwarzschild BH \cite{nonex1}.
The reason is clear, for both addressed cases, Eq.~(\ref{tem})
reduces to the ordinary temperature of Schwarzschild BH
\cite{Bohr3}. It is also worthwhile mentioning some similarities and
differences between the properties of an excited Schwarzschild BH in
the Bekenstein and R\'{e}nyi entropy formalisms. $i$) Both the
Bekenstein~(\ref{0ent}) and R\'{e}nyi~(\ref{ent0}) entropies are
vanished for $n=n_{max}$, $ii$) in the framework of the Bekenstein
entropy, temperature is always independent of $n$, and $iii$) heat capacity is always negative, independent of the value of $n$, in
the regime of the Bekenstein entropy (the $\beta=0$ limit
of~(\ref{c})), while $C$ can be positive depending on the values of
$n$ and $\beta$. For example, whereas $n=n_{max}$, we have $C>0$ if
$\beta>\beta_0$.
\section{Summary and Concluding Remarks}

The R\'{e}nyi entropy for a BH has been obtained through EUP. By
using the recent Bohr-like approach to BH quantum physics, we have
also shown that, for excited BHs, the R\'{e}nyi entropy is a
function of the BH principal quantum number, i.e. the BH quantum
excited state. Some thermodynamic properties of excited BH meeting
R\'{e}nyi entropy have also been addressed. The results show that
although there are two phases at $n\rightarrow n_{max}$, only the
phase with $\beta>\beta_0$ can be stable. Moreover, whenever
$n\rightarrow n_{max}$, then entropy is vanished, and temperature
takes its non-zero possible minimum value ($T_B$) if
$\beta=\beta_0$.

Evaporation time of the excited BHs has also been studied. Although
we addressed the $n=n_{max}$ case in our study, in reality, the
maximum possible value of $n$, namely $N_{max}$, may be limited as
$N_{max}=n_{max}-1$ whenever the Planck mass and the Planck distance are
approached \cite{33,Bohr3}. In this manner, non of the Bekenstein
and R\'{e}nyi entropies are vanished when the maximum value of $n$
is taken by the excited BH. It leads to the interesting results in
the R\'{e}nyi entropy framework, in agreement with the third law of
thermodynamics, $i$) based on Eq.~(\ref{ent0}), entropy takes its
minimum value which is non-zero, and $ii$) from Eq.~(\ref{tem}), the
minimum of temperature can be positive or even greater than $T_B$
depending on the value of $\beta$.
\section*{Acknowledgment}
The work of H.
Moradpour has been supported financially by Research Institute for
Astronomy \& Astrophysics of Maragha (RIAAM).

\end{document}